\documentclass{PoS}

\usepackage{amsmath,relsize}

\newcommand{\EQ}[1]{\begin{equation}\begin{split} #1
\end{split}\end{equation}}
\newcommand{\SP}[1]{\begin{equation}\begin{split} #1
\end{split}\end{equation}}

\def\Tr{{\rm Tr}}
\def\tr{{\rm tr}}

\newcommand*{\Dsl}[0]{{\rlap{\kern2.25pt /}{D}}}

\title{Investigating corrections to a Gaussian distribution of the complex phase}

\ShortTitle{Investigating corrections to a Gaussian distribution}

\author{Jeff Greensite\\
        $^1$ Physics and Astronomy Department, San Francisco State University, 1600 Holloway Ave., San Francisco CA, 94132 USA\\
        $^2$ Niels Bohr International Academy, Blegdamsvej 17, 2100 Copenhagen \O, Denmark\\
        E-mail: \email{greensit@sfsu.edu}}

\author{\speaker{Joyce C. Myers}\\
        Discovery Center, The Niels Bohr Institute, University of Copenhagen, Blegdamsvej 17, 2100 Copenhagen \O, Denmark\\
        E-mail: \email{jcmyers@nbi.dk}}
        
\author{K. Splittorff\\
        Discovery Center, The Niels Bohr Institute, University of Copenhagen, Blegdamsvej 17, 2100 Copenhagen \O, Denmark\\
        E-mail: \email{split@nbi.dk}}

\abstract{It has been suggested that the density of states approach to performing lattice simulations in QCD with nonzero chemical potential can be modified to improve the signal to noise ratio by performing a cumulant expansion of the complex phase of the fermion determinant, and then simplified by truncating the expansion after the first non-zero cumulant. This truncation corresponds to approximating the distribution of the complex phase of the fermion determinant by a Gaussian form. The crucial question is: how large are the other cumulants? We calculate the distribution of the complex phase from the hadron resonance gas model and from a combined lattice strong coupling and hopping expansion. In the case of the hadron resonance gas model the distribution takes a Gaussian form, but from the strong coupling and hopping expansion there are corrections. We discuss the implications to lattice simulations.}

\FullConference{31st International Symposium on Lattice Field Theory - LATTICE 2013\\
		July 29 - August 3, 2013\\
		Mainz, Germany}

\begin{document}

\section{Introduction}

Calculating the phase diagram of QCD at non-zero temperature and chemical potential is one of the most enduring unsolved problems that faces the particle physics community. The challenges lie on the theoretical side as well as the experimental. On the experimental side it is possible to use heavy ion collisions to locate critical surfaces in the QCD phase diagram. In a system at equilibrium the correlation length is expected to diverge at a transition point, a signiture of which should be visible from fluctuations of thermodynamic observables such as the baryon number, electric charge, and strangeness. Significant progress has been made at small chemical potentials \cite{Ejiri:2005wq}, but it is difficult to interpret results at larger chemical potentials where there are no reliable theoretical models available \cite{Adamczyk:2013dal}. On the theoretical side the most natural way to calculate at finite chemical potential would seem to involve the use of lattice simulations, since these have been effective in obtaining the phase diagram at zero chemical potential. However, at non-zero chemical potential there is the notorious ``sign problem", which results because the fermion determinant becomes complex,
\EQ{
\det(\Dsl+\gamma_0 \mu+m) = \left| \det(\Dsl+\gamma_0 \mu+m) \right| e^{i \theta} \, ,
}
making the use of conventional Monte Carlo methods based on importance sampling ineffective.

Over the years there have been several proposals for simulating systems with a sign problem which involve adapting techniques to manage a complex fermion determinant, or converting the system to one in which the fermion determinant becomes real. For a review, see for example \cite{deForcrand:2010ys,Aarts:2013bla,Splittorff:2006vj}. The procedure which we will discuss here, the density of states method \cite{Gocksch:1988iz,Anagnostopoulos:2001yb,Fodor:2007vv,Ejiri:2007ga,Fukushima:2010bq}, specifically when used in combination with a cumulant expansion of the complex phase \cite{Ejiri:2007ga,Nakagawa:2011eu,Saito:2012nt,Saito:2013vja}, is in some sense, a combination of these ideas. First, the density of states method is a form of re-weighting, in which the complex phase factor of the fermion determinant becomes part of the observable, rather than part of the action,
\EQ{
\langle {\cal O} \rangle_{QCD} = \frac{\langle {\cal O} e^{i N_f \theta} \rangle_{pq}}{\langle e^{i N_f \theta} \rangle_{pq}} \, ,
\label{obs}
}
where $pq$ indicates that the expectation value is with respect to the phase-quenched theory in which $\det(\Dsl+\gamma_0 \mu+m)$ is replaced by its absolute value $\left| \det(\Dsl+\gamma_0 \mu + m) \right|$. The density of states method relies on the introduction of a density of some fixed quantity $X$,
\EQ{
\rho_{pq}(X) \equiv \langle \delta(X-X') \rangle_{pq} = \frac{1}{Z_{pq}} \int {\rm D}U~ \delta(X - X'(U)) \left| \det (\Dsl+\gamma_0 \mu+m) \right|^{N_f} e^{-S_g} \, ,
}
such that observables can be obtained using reweighting from
\EQ{
\langle {\cal O} \rangle_{QCD} = \frac{\mathop{\mathlarger {\int}} {\rm d}X \left[ \dfrac{\langle \delta(X-X') {\cal O} e^{i N_f \theta} \rangle_{pq}}{\rho_{pq}(X)} \right] \rho_{pq}(X)}{\mathop{\mathlarger {\int}} {\rm d}X \left[ \dfrac{\langle \delta(X-X') e^{i N_f \theta} \rangle_{pq}}{\rho_{pq}(X)} \right] \rho_{pq}(X)} \, ,
}
where the distribution is normalized as
\EQ{
\int {\rm d}X \rho_{pq}(X) = 1 \, .
}
This implies that the density, or distribution, has a direct probability interpretation.

The difficulty in calculating expectation values using re-weighting is that in the large volume limit \cite{Splittorff:2006fu}
\EQ{
\langle e^{i \theta} \rangle_{pq} = e^{-c V} \, .
\label{noise}
}
This implies that exponential accuracy is needed in (\ref{obs}) in order to avoid a signal-to-noise problem. The issue can potentially be resolved through the use of the cumulant expansion \cite{Ejiri:2009hq}
\EQ{
\langle e^{i \theta} \rangle_{X} = \exp \left[ -\frac{1}{2} \langle \theta^2 \rangle_c + \frac{1}{4!} \langle \theta^4 \rangle_c - ... \right] \, ,
\label{cumu-exp}
}
with
\SP{
\langle \theta^2 \rangle_c &= \langle \theta^2 \rangle_X \, , \\
\langle \theta^4 \rangle_c &= \langle \theta^4 \rangle_X - 3 \langle \theta^2 \rangle_X^2 \, , \\
&... \, ,
\label{cumulants}
}
where\footnote{In what follows we will take $X = \theta$, whereas in simulations one takes $X$ as the average plaquette, or Polyakov line, or sometimes multiple observables are kept fixed. In our analysis we make some comparisons with simulation results but they will be qualitative. Also, we do not calculate $\langle {\cal O} \rangle_X$, but rather $\langle {\cal O} \rangle_{pq} = \int {\rm d}X \rho_{pq}(X) \langle {\cal O} \rangle_X$.}
\EQ{
\langle {\cal O} \rangle_X = \frac{\langle \delta(X'-X) {\cal O} \rangle_{pq}}{\rho_{pq} (X)} \, .
}
The quantities in the expectation values on the r.h.s. of (\ref{cumu-exp}) are real and positive so it may appear that the noise problem of (\ref{noise}) has been resolved. Furthermore, the technique of WHOT-QCD is to approximate the expansion by the first cumulant $\langle \theta^2 \rangle_c$ \cite{Ejiri:2007ga}. This corresponds to a Gaussian distribution of the complex phase of the fermion determinant and is motivated by the central limit theorem. But has the sign problem been resolved, or has it relocated into the higher order cumulants? How large are $\langle \theta^4 \rangle_c$, $\langle \theta^6 \rangle_c$, ... and what would be the consequences if they were comparable to $\langle \theta^2 \rangle_c$?

Perhaps the most effective way to determine if the cumulant expansion converges is to measure the cumulants directly, as members of WHOT-QCD have been careful to do in their simulations \cite{Ejiri:2007ga,Nakagawa:2011eu,Saito:2012nt,Saito:2013vja}, but it is difficult to know if/when enough cumulants have been computed, and to determine how large higher order cumulants are, since the statistics needed to accurately obtain them is higher. Here we explore two ways to calculate the higher order cumulants analytically. First, we consider the hadron resonance gas model \cite{Dashen:1969ep} including contributions from ground state mesons of spin $0$ and $1$, and baryons of spin $\frac{1}{2}$ and $\frac{3}{2}$. Second, we calculate them from a combined lattice strong coupling and hopping expansion \cite{Langelage:2010yr,Fromm:2011qi}. The details of these calculations can be found in \cite{Greensite:2013gya}, and some consequences of the results are discussed in \cite{Greensite:2013vza}.

We begin with a statement of our results from the hadron resonance gas model and the combined lattice strong coupling and hopping expansions and summarize the immediate consequences. A summary of the calculation of the hadron resonance gas model is presented in Section \ref{hrg}, and a summary of the strong coupling expansion in Section \ref{strong}. Full details of these calculations can be found in \cite{Greensite:2013gya}.

\section{Moments of the complex phase}

The distribution of the complex phase of the fermion determinant can be calculated analytically by means of the Fourier transform
\EQ{
\rho(\theta) = \langle \delta(\theta - \theta') \rangle = 2 \int_{-\infty}^{\infty} \frac{{\rm d}p}{2\pi} e^{-2 i p \theta} \langle e^{2 i p \theta'} \rangle \, .
\label{fourier}
}
We just need to calculate the moments \cite{Lombardo:2009aw,Splittorff:2007zh}
\EQ{
\langle e^{2 i p \theta'} \rangle = \frac{Z_{YM}}{Z} \bigg{\langle} \frac{\det^p(\Dsl+\gamma_0 \mu+m)}{\det^p(\Dsl-\gamma_0 \mu+m)} {\det}^{N_f}(\Dsl+\gamma_0 \mu+m) \bigg{\rangle}_{YM} \, .
\label{full-moms}
}
Since these moments take the form of a partition function, it is natural that our results can be expressed as an exponential of a quantity proportional to the $3$-volume $V$,
\EQ{
\langle e^{2 i p \theta'} \rangle = \exp \left[ - f(p) V \right] \, ,
\label{mom-exp}
}
where $f(p)$ is the $3$-volume-independent free energy density over temperature. For the hadron resonance gas we find that the phase-quenched moments take the form \footnote{In practice we calculate the expectation value $\langle e^{2 i p \theta'} \rangle$ in the full theory, rather than the phase-quenched theory, since the expectation value in the full theory has the more straightforward definition in (\ref{full-moms}). It is possible to convert between the two using $\langle e^{2 i p \theta'} \rangle = \langle e^{2 i q \theta'} \rangle_{pq}$ with $p = q - \frac{N_f}{2}$. To convert between the distributions one can use $\rho(\theta) = \frac{Z_{pq}}{Z} e^{i N_f \theta} \rho_{pq}(\theta)$ \cite{Splittorff:2007zh}.}
\EQ{
\langle e^{2 i p \theta'} \rangle_{pq} = \exp[-p^2 x_1] \, ,
\label{hrg-moms}
}
which leads to a Gaussian distribution via (\ref{fourier}),
\EQ{
\rho_{pq}(\theta) = 2 \int_{-\infty}^{\infty} \frac{{\rm d}p}{2\pi} e^{-2 i p \theta} e^{-p^2 x_1} = \frac{1}{\sqrt{\pi x_1}} e^{-\theta^2 / x_1} \, .
}
For the combined strong coupling and hopping expansions, working at ${\cal O}(\beta^{N_t})$ and in the confined phase, the moments take the form
\EQ{
\langle e^{2 i p \theta'} \rangle_{pq} = \exp[-p^2 x_1 - p^4 x_2 - p^6 x_3 - ...] \, ,
\label{gen-moms}
}
which results in a distribution with corrections to a Gaussian form
\EQ{
\rho_{pq}(\theta) = 2 \int_{-\infty}^{\infty} \frac{{\rm d}p}{2\pi} e^{-2 i p \theta} e^{-p^2 x_1 -p^4 x_2 - p^6 x_3 - ...} \, ,
\label{gen-dist}
}
where all of the $x_n$ are ${\cal O}(V)$. The corrections resulting in the strong coupling and hopping expansion are a consequence of working to sufficiently high order in $\beta$ and $\frac{1}{m a}$, such that at least $6$ Polyakov lines are available to form color singlets. Specifically, we find the corrections to be a consequence of working at finite $N_c$. The corrections in (\ref{gen-moms}) can only appear when an even number of Polyakov lines can be combined in such a way that there is a nonzero contribution for $N_c = 3$, which vanishes for $N_c = \infty$. For the strong coupling and hopping expansion, expectation values formed with $6$ Polaykov lines occur at ${\cal O}(h^4)$ when working at ${\cal O}(\beta^{N_t})$, and at ${\cal O}(h^6)$ when working at ${\cal O}(\beta^{0})$. In the hadron resonance gas model one combines quark lines into color singlets in place of Polyakov lines. Since we only consider $2$-quark combinations (mesons), and $3$-quark combinations (baryons), in the non-interacting limit, higher order contributions to the distribution do not appear.

At this point it is helpful to notice that the $x_n$ are related to the cumulants in a simple way. A cumulant expansion of the moments takes the form
\EQ{
\log \langle e^{2 i p \theta} \rangle_{pq} = \sum_{n=1}^{\infty} \frac{(2 i p)^n}{n!} \langle \theta^n \rangle_c \, .
}
Plugging in the most general expression for the moments in (\ref{gen-moms}) reveals that each cumulant corresponds to one of the $x_n$,
\EQ{
x_n = - \frac{(2 i)^{2 n}}{(2 n)!} \langle \theta^{2 n} \rangle_c
\label{xn}
}
such that
\EQ{
- \log \langle e^{2 i \theta} \rangle_{pq} = x_1 + x_2 + ... \, .
}
It is clearly necessary to have $x_1 \gg x_2, x_3, x_4, ...$ for the Gaussian approximation to succeed, but they are all ${\cal O}(V)$ so this is not guaranteed. In addition, there is an issue which will come up when calculating them in simulations from the $\langle \theta^n \rangle_{pq}$. Notice that $\langle \theta^n \rangle_{pq}$ can be obtained, using (\ref{gen-moms}), from
\EQ{
\langle \theta^n \rangle_{pq} = \left[ \left( \frac{1}{2 i} \right)^n \frac{{\rm d}}{{\rm d}q^n} e^{-q^2 x_1 - q^4 x_2 - ...} \right]_{q=0} \, ,
}
where it is clear that since the $x_n$ are ${\cal O}\left( V \right)$, then the $\langle \theta^{2 n} \rangle_{pq}$ are ${\cal O} \left( V^n \right)$. In simulations the cumulants $\langle \theta^{2 n} \rangle_c$ are obtained from calculations of the $\langle \theta^{2 n} \rangle_{pq}$ using (\ref{cumulants}). Therefore, in order to obtain an ${\cal O}(V)$ result for the higher order cumulant $\langle \theta^{2 n} \rangle_c$ with $n > 1$, it must happen that there are cancellations of the ${\cal O}(V^n)$ contributions.

An important point to make before we move on is that our calculations are valid in the confined phase for chemical potentials $\mu < \frac{m_{\pi}}{2}$. For larger chemical potentials the distribution is expected to take a Lorentzian form \cite{Lombardo:2009aw}, where the authors obtained the distribution from chiral perturbation theory and one-dimensional QCD. In order to extend our calculations into the region of $\mu > \frac{m_{\pi}}{2}$ it would be necessary to consider a background with a condensate of bound states containing one quark with chemical potential $\mu$, and one with chemical potential $-\mu$.

\section{Hadron resonance gas}
\label{hrg}

The hadron resonance gas model \cite{Dashen:1969ep} provides a form of the partition function for free hadrons: including baryons and mesons in their ground states and their resonances. It provides an effective description of net baryon number, electric charge, and strangeness fluctuations, for example, obtained from particle abundances in heavy-ion collisions \cite{Andronic:2005yp,Karsch:2010ck,Bazavov:2013dta}, for sufficiently small temperatures and chemical potentials.

To obtain the result for the expectation value $\langle e^{2 i p \theta'} \rangle_{pq}$ in (\ref{hrg-moms}) for the hadron resonance gas, it is useful to notice that the general form of $\langle e^{2 i p \theta'} \rangle$ in (\ref{full-moms}) corresponds to a partition function of a theory with $p+N_f$ quarks from the determinants in the numerator, and $p$ "ghost quarks" from the determinant in the denominator. Therefore, to obtain the expectation value it is necessary to compute the spectrum of the quarks and ghost quarks, which proceeds in an analogous way to calculating the hadron spectrum of the standard model. For all of the details see \cite{Greensite:2013gya}.

Our calculation of the moments $\langle e^{2 i p \theta'} \rangle$ includes all possible spectral combinations of mesons with spin $s = 0$, $1$, and baryons with spin $\frac{1}{2}$, $\frac{3}{2}$, for $2 p + N_f$ flavors. The precise contributions are obtained by performing the decompositions, from $SU(2(2p+N_f))$ to $SU(2p+N_f)_{flavor} \times SU(2)_{spin}$. For baryons the relevant decomposition is obtained from
\EQ{
{\bf n} \otimes {\bf n} \otimes {\bf n} = \left( {\bf \frac{n(n+1)(n+2)}{6}} \right) \oplus ... \, ,
}
\EQ{
\left( {\bf \frac{n(n+1)(n+2)}{6}} \right) \rightarrow \left( {\bf \frac{\frac{n}{2}(\frac{n}{2} - 1)(\frac{n}{2} + 1)}{3}} \right)_{{\bf 2}} \oplus \left( {\bf \frac{\frac{n}{2}(\frac{n}{2}+1)(\frac{n}{2}+2)}{6}} \right)_{{\bf 4}} \, ,
}
where the arrow indicates the decomposition $SU(n) \rightarrow SU(\frac{n}{2}) \times SU(2)$, and ${\bf R}_{{\bf g}}$ is the decomposed product with ${\bf R} \in SU(\frac{n}{2})$ and ${\bf g} \in SU(2)$. We note that for ground state baryons it is necessary that the total wavefunction is completely antisymmetric so in this case one only needs to consider the decomposition of the symmetric representation in $SU(2(2p+N_f))$ (flavor and spin combined) since the wavefunction is antisymmetric in color. For mesons the decomposition is
\EQ{
{\bf n} \otimes {\bf \bar{n}} \rightarrow \left[ {\bf \left( \frac{n}{2} \right)^2 - 1} \right]_{\bf {3}} \oplus \left[ {\bf \left( \frac{n}{2} \right)^2 - 1} \right]_{{\bf 1}} \oplus {\bf 1_3} \oplus {\bf 1_1} \, .
}

To obtain $\langle e^{2 i p \theta'} \rangle$ for a free hadron gas one simply needs to add up the free energies from all possible hadronic states. For free mesons the free energy is
\EQ{
F_{g}^{M}(\mu_I) = -g \frac{m_M^2 T^2}{\pi^2} \sum_{n=1}^{\infty} \frac{1}{n^2} {\bf K}_2 (n m_M / T) \cosh[2 n I_3 \mu_I / T] \, ,
}
where $g = 2 s + 1$ is the spin degeneracy, $I_3 \equiv \frac{1}{2} \left[ (N_u - N_{{\bar u}}) - (N_d - N_{{\bar d}}) \right]$ is the third isospin component, and $\mu_I = \frac{1}{2}(\mu_u - \mu_d)$ is the isospin chemical potential. For free baryons the free energy is
\EQ{
F_g^{B}(\mu_B - 2 I_3 \mu_I) = g \frac{m_B^2 T^2}{\pi^2} \sum_{n=1}^{\infty} \frac{(-1)^n}{n^2} {\rm K}_2(n m_B/T) \cosh[(\mu_B - 2 I_3 \mu_I) n \beta] \, ,
}
where $\mu_B = \mu_u + \mu_d + ...$ is the baryon chemical potential. We are working in the approximation that the $N_f$ flavors are degenerate, each with a quark chemical potential $\mu_q = \mu$.

Noting that each ghost quark contributes a factor of $-1$ to the free energy, and $-\mu$ to the chemical potential, the result is
\EQ{
\langle e^{2 i p \theta'} \rangle_{pq} = e^{- p^2 x_1}
}
with
\EQ{
x_1 = F^M(2 \mu) - F^M(0)+F^B(3\mu)-F^B(0),
}
where
\SP{
F^M(\mu) &\equiv k_M [F_1^M(\mu) + F_3^M(\mu)] \, , \\
F^B(\mu) &\equiv k_B [2 F_2^B(\mu) + F_4^B(\mu)] \, .
}
Since $x_n = 0$ for $n > 1$ the higher order cumulants $\langle \theta^{2n} \rangle_c$ for $n > 1$ are zero and the distribution takes a Gaussian form.

\section{Taylor expansion}

Before moving on to the strong coupling calculation it is possible to show that it is in principle natural to have nonzero higher order cumulants. This can be seen by performing a Taylor expansion of $\log \langle e^{2 i p \theta'} \rangle$ around $\mu/T = 0$. Defining $M(\mu) \equiv \det(\Dsl + \gamma_0 \mu + m)$ and $D^{(n)}(\mu) \equiv \frac{\partial^n}{\partial(\mu/T)^n} \frac{M(\mu)^{p+N_f}}{M(-\mu)^p}$,
\SP{
&\log\left[ \frac{Z_{YM}}{Z} \bigg{\langle} \frac{M(\mu)^{p+N_f}}{M(-\mu)^p} \bigg{\rangle}_{YM} \right] = \frac{1}{2!} \left( \frac{\mu}{T} \right)^2 \left[ \frac{\langle D^{(2)}(0) \rangle_{YM}}{\langle M(0)^{N_f} \rangle_{YM}} \right] \\
&\hspace{2cm}+ \frac{1}{4!} \left( \frac{\mu}{T} \right)^{4} \left[ \frac{\langle D^{(4)}(0) \rangle_{YM}}{\langle M(0)^{N_f} \rangle_{YM}} - 3 \frac{\langle D^{(2)}(0)\rangle_{YM}^2}{\langle M(0)^{N_f} \rangle_{YM}^2} \right] \\
&\hspace{2cm}+ \frac{1}{6!} \left( \frac{\mu}{T} \right)^{6} \left[ \frac{\langle D^{(6)}(0) \rangle_{YM}}{\langle M(0)^{N_f} \rangle_{YM}} - 15 \frac{\langle D^{(2)}(0) \rangle_{YM} \langle D^{(4)}(0) \rangle_{YM}}{\langle M(0)^{N_f} \rangle_{YM}^2} + 30 \frac{\langle D^{(2)}(0) \rangle_{YM}^3}{\langle M(0)^{N_f} \rangle_{YM}^3} \right] \\
&\hspace{2cm}+ {\cal O}\left( \frac{\mu}{T} \right)^8 - \log \left[ \frac{Z}{Z_{YM}} \right] \, .
}
Evaluating the derivatives and collecting terms with like powers of $p$ results in a series of special relationships which must hold to make $x_2, x_3, ... = 0$. For example, to make $x_2 = 0$ at ${\cal O}\left(\frac{\mu}{T}\right)^4$ it is required that
\EQ{
\langle M(0)^{N_f} \rangle_{YM} \langle M(0)^{N_f - 4} M'(0)^4 \rangle_{YM} = 3 \langle M(0)^{N_f - 2} M'(0)^2 \rangle_{YM}^2 \, .
}
There are similar relationships which must hold at higher orders in $\frac{\mu}{T}$ and additional relationships that must hold at ${\cal O}\left( \frac{\mu}{T} \right)^4$ which can be determined by collecting terms with higher powers of $p$.

\section{Lattice strong coupling and hopping expansion}
\label{strong}

The Taylor expansion of the complex phase moments above indicates that higher order cumulants will appear in the absence of special relationships at zero chemical potential. It is possible to show analytically that these higher order cumulants are indeed realized by means of a combined lattice strong coupling and hopping expansion.

The combined use of the lattice strong coupling and hopping expansions allows for analytical calculations by way of an effective Polyakov line action. Recently this technique has been used successfully to obtain information about the phase diagram of QCD with a chemical potential \cite{Fromm:2012uj,Fromm:2012eb,Fromm:2011aa,Fromm:2011qi}. In what follows we will use the effective Polyakov line action to calculate the $x_n$ from the phase angle moments $\langle e^{2 i p \theta} \rangle$ in (\ref{full-moms}).

\subsection{Lattice strong coupling expansion}

One of the simplifications of working at strong coupling is that the effective action can be formulated as a function of Polyakov lines. After integrating out the spatial link variables the lattice Yang-Mills partition function can be further simplified by means of the character expansion \cite{Drouffe:1983fv,Green:1983sd,Montvay:1994cy}
\EQ{
Z_{YM} = \int_{SU(N_c)} \prod_{{\bf z}} {\rm d}W_{{\bf z}} \prod_{\langle {\bf x} {\bf y} \rangle} \left[ 1 + \sum_{R} \lambda_R \left[ \chi_R(W_{{\bf x}}) \chi_{R}(W_{{\bf y}}^{\dagger}) + \chi_R(W_{{\bf x}}^{\dagger}) \chi_R(W_{{\bf y}}) \right] \right] \, ,
}
where $\chi_R(W_{{\bf x}}) = \Tr_R(W_{\bf x})$ are the characters of the Polyakov lines $W_{{\bf x}} = \prod_{\tau = 0}^{N_{\tau - 1}} U_0({\bf x}, \tau)$, $\prod_{\langle {\bf x} {\bf y} \rangle}$ is over nearest neighbor sites, and the $\lambda_R$ are expansion parameters in powers of $\frac{1}{g^2 N_c}$.

Working at leading order corresponds to truncating the sum over $R$ at the fundamental representation, such that
\EQ{
e^{-S_{YM}} \rightarrow 1 + \lambda_1 \sum_{\langle {\bf x} {\bf y} \rangle} \left[ \tr(W_{{\bf x}}) \tr(W_{{\bf y}}) + \tr(W^{\dagger}_{{\bf x}}) \tr(W_{{\bf y}}) \right] \, ,
}
with $\lambda_1 = \left( \frac{1}{g^2 N_c} \right)^{N_{\tau}}$. This is the limit we consider from here on.

\subsection{Hopping expansion}

The fermion determinant can be expanded in the static, heavy quark limit using the hopping expansion \cite{Fromm:2011qi} (see also \cite{DePietri:2007ak})
\EQ{
\log \det \left( \Dsl + \gamma_0 \mu + m \right) = a_1 h \left[ e^{\mu/T} \tr W_{{\bf x}} + e^{-\mu/T} \tr W_{{\bf x}}^{\dagger} \right] + a_2 h^2 \left[ e^{2 \mu/T} \tr(W_{{\bf x}}^2) + e^{-2\mu/T} \tr(W_{{\bf x}}^{\dagger}) \right] + ... \, .
\label{ferm-det}
}
For Wilson fermions
\EQ{
a_n = 2 \frac{(-1)^n}{n} \, , \hspace{1cm} h = \left( 2 \kappa_f \right)^{N_{\tau}} \, , \hspace{1cm} \kappa_f = \frac{1}{2(m a+d+1)} \, .
}
By calculating the moments $\langle e^{2 i p \theta'} \rangle$, we obtain the leading order contributions to the cumulants $x_n$, which are at least ${\cal O}(h^{2 n})$. At ${\cal O}(\lambda_1^0)$ we calculate the leading order contribution to $x_1, ..., x_6$. At ${\cal O}(\lambda_1)$ we calculate the leading order contribution to $x_1, x_2, x_3$. In both cases the calculations are carried out in the confined phase.

To obtain $\langle e^{2 i p \theta'} \rangle$ from (\ref{full-moms}) in the heavy quark limit we expand in the hopping parameter $h$,
\EQ{
Q \equiv \bigg{\langle} \frac{\det^p(\Dsl+\gamma_0 \mu+m)}{\det^p(\Dsl-\gamma_0 \mu+m)} {\det}^{N_f}(\Dsl+\gamma_0 \mu+m) \bigg{\rangle}_{YM} = 1 + q_1 h + q_2 h^2 + ... \, .
\label{Q}
}
Using (\ref{ferm-det}), the contributions up to ${\cal O}(h^2)$ take the form
\EQ{
q_1 = 2 a_1 N_f \cosh(\mu/T) \sum_{{\bf x}} \langle \tr W_{{\bf x}} \rangle \, ,
}
\SP{
q_2 = \,\, &2 a_1^2 p(p+N_f) \left[ \cosh(2 \mu/T) - 1 \right] \sum_{{\bf x},{\bf y}} \left[ \langle \tr W_{{\bf x}} \tr W_{{\bf y}} \rangle - \langle \tr W_{{\bf x}} \tr W_{{\bf y}}^{\dag} \rangle \right] \\
&+ 2 a_2 N_f \cosh(2\mu/T) \sum_{{\bf x}} \langle \tr (W_{{\bf x}}^2) \rangle + a_1^2 N_f^2 \sum_{{\bf x},{\bf y}} \left[ \cosh(2 \mu/T) \langle \tr W_{{\bf x}} \tr W_{{\bf y}} \rangle + \langle \tr W_{{\bf x}} \tr W_{{\bf y}}^{\dag} \rangle \right] \, .
}
where we used the fact that the YM vacuum is charge conjugation symmetric, such that \\$\langle [\tr (W_{{\bf x}}^n)]^i ...  [\tr (W_{{\bf y}}^{\dag m})]^j \rangle_{YM} = \langle [\tr (W_{{\bf x}}^{\dag n})]^i ...  [\tr (W_{{\bf y}}^{m})]^j \rangle_{YM}$. The overall result for (\ref{Q}) exponentiates as in (\ref{mom-exp}) (we checked this to ${\cal O}(h^4)$) so it is sufficient to consider the ${\cal O}(V)$ terms to find the contribution to $x_1$ defined in (\ref{gen-moms}). The result is
\EQ{
x_1 = 2 a_1^2 h^2 \left[ \cosh(2 \mu/T) - 1 \right] \sum_{{\bf x},{\bf y}} \left[ \langle \tr W_{{\bf x}} \tr W_{{\bf y}}^{\dag} \rangle - \langle \tr W_{{\bf x}} \tr W_{{\bf y}} \rangle \right] + {\cal O}(h^3) \, .
}
While it is clear that there are no contributions to the $x_n$ for $n > 1$ at this order, since there are no terms with ${\cal O}(p^{\nu})$ with ${\nu} > 2$, they do begin to appear at ${\cal O}(h^4)$. From here on we work in the confined phase. The above result simplifies since $\langle \tr W_{{\bf x}} \rangle = \langle \tr W_{{\bf x}} \tr W_{{\bf y}} \rangle = 0$, and $\langle \tr W_{{\bf x}} \tr W_{{\bf x}}^{\dag} \rangle = N_s$, where $N_s$ is the number of spatial lattice sites.

Calculating the higher order contributions to (\ref{Q}) amounts to calculating expectation values of Polyakov lines. We define $P_n = \tr(W_{{\bf x}}^n)$, $P_n^* = \tr(W_{{\bf x}}^{\dagger n})$. At each order, all contributions which result in color singlets must be obtained. For example,
\SP{
\langle P_1 P_1^* \rangle_{YM} &= \text{singlets in}~ {\bf 3} \otimes {\bf {\bar 3}} = 1 \, , \\
\langle P_1^2 P_1^{* 2} \rangle_{YM} &= \text{singlets in}~ {\bf 3} \otimes {\bf 3} \otimes {\bf {\bar 3}} \otimes {\bf {\bar 3}} = 2 \, , \\
\langle P_1^3 \rangle_{YM} &= \text{singlets in}~ {\bf 3} \otimes {\bf 3} \otimes {\bf 3} = 1 \, \\
\langle P_1^4 P_1^* \rangle_{YM} &= \text{singlets in}~ {\bf 3} \otimes {\bf 3} \otimes {\bf 3} \otimes {\bf 3} \otimes {\bf {\bar 3}} = 3 \, , \\
\langle P_2 P_1 \rangle_{YM} &= \langle (P_1^2 - 2 P_1^*) P_1 \rangle_{YM} = -1 \, , \\
...&
}
Note that the third and fourth vevs only contribute for $SU(3)$, and that the last is $-2$ when $N_c = \infty$. In general the nonzero expectation values take the form
\EQ{
\int_{SU(N_c)} {\rm d}W (\tr W \tr W^{\dagger})^l (\tr W)^{N_c m} (\tr W^{\dagger})^{N_c n} \ne 0 \, ,
}
where $l, m, n = 0, 1, 2, ...$.

Our results for the leading order contributions to the cumulants from the hopping expansion, working at ${\cal O}(\lambda_1^0)$, are worked out in detail in \cite{Greensite:2013gya} and summarized in Table \ref{tab1}. Our results at ${\cal O}(\lambda_1)$ are summarized in Table \ref{tab2}. For $SU(3)$, contributions resulting from $x_n \ne 0$ for $n>1$ imply that there are non-zero higher order cumulants (\ref{xn}). Moreover, the trend appears to be that they become more significant with increasing $\mu/T$ or $\beta$, or decreasing $m$. It is interesting to observe that in the limit $N_c \rightarrow \infty$ the corrections vanish.

\begin{table}[h]
\begin{tabular}{|l|l|l|}
\hline
& $N_c = 3$ & $N_c = \infty$\\
\hline
$x_1$ & $4 a_1^2 {\color[rgb]{0.8,0,0}h^2} \sinh^2(\mu/T) + {\cal O}(h^3)$ & $4 a_1^2 {\color[rgb]{0.8,0,0}h^2} \sinh^2(\mu/T) + {\cal O}(h^3)$\\
\hline
$x_2$ & $4 {\color[rgb]{0.8,0,0}h^5} \sinh^4(\mu/T) \cosh(\mu/T) \left[ 8 a_1^3 a_2 - a_1^5 N_f \right] N_s + {\cal O}(h^6)$ & $0 + {\cal O}(h^6)$\\
\hline
$x_3$ & $- \frac{8}{9} N_s a_1^6 {\color[rgb]{0.8,0,0}h^6} \sinh^6(\mu/T) + {\cal O}(h^7)$ & $0 + {\cal O}(h^7)$\\
\hline
$x_4$ & $- \frac{44}{45} N_s a_1^8 {\color[rgb]{0.8,0,0}h^8} \sinh^8(\mu/T) + {\cal O}(h^9)$ & $0 + {\cal O}(h^9)$\\
\hline
$x_5$ & $- \frac{112}{225} N_s a_1^{10} {\color[rgb]{0.8,0,0}h^{10}} \sinh^{10}(\mu/T) + {\cal O}(h^{11})$ & $0 + {\cal O}(h^{11})$\\
\hline
$x_6$ & $\frac{3488}{14175} N_s a_1^{12} {\color[rgb]{0.8,0,0}h^{12}} \sinh^{12}(\mu/T) + {\cal O}(h^{13})$ & $0 + {\cal O}(h^{13})$\\
\hline
\end{tabular}
\caption{Leading order contributions to the cumulants at ${\cal O}(\lambda_{1}^{0})$}
\label{tab1}
\end{table}

\begin{table}[h]
\begin{tabular}{|l|l|l|}
\hline
& $N_c = 3$ & $N_c = \infty$\\
\hline
$x_1$ & $0 + {\cal O}(h^3)$ & $0 + {\cal O}(h^3)$\\
\hline
$x_2$ & $-24 \lambda_1 N_s a_1^4 {\color[rgb]{0.8,0,0}h^4} \sinh^4(\mu/T) + {\cal O}(h^5)$ & $0 + {\cal O}(h^5)$\\
\hline
$x_3$ & $- 80 \lambda_1 N_s a_1^6 {\color[rgb]{0.8,0,0}h^6} \sinh^6(\mu/T) + {\cal O}(h^7)$ & $0 + {\cal O}(h^7)$\\
\hline
\end{tabular}
\caption{Leading order contributions to the cumulants at ${\cal O}(\lambda_1)$}
\label{tab2}
\end{table}

\subsection{Cumulants}

\begin{figure}
\begin{minipage}{0.48\textwidth}
\includegraphics[width=0.96\textwidth]{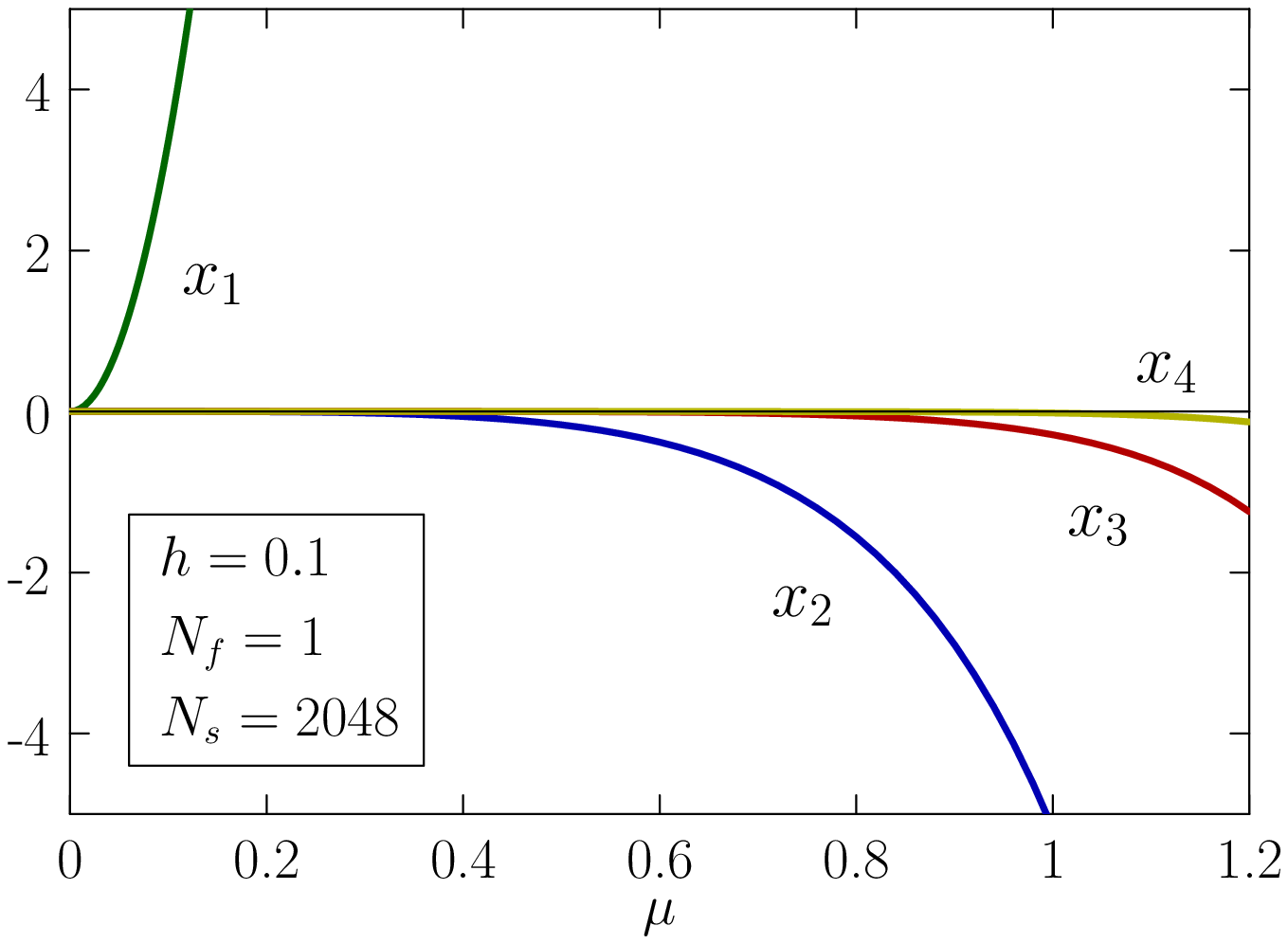}
\end{minipage}
\begin{minipage}{0.48\textwidth}
\includegraphics[width=0.96\textwidth]{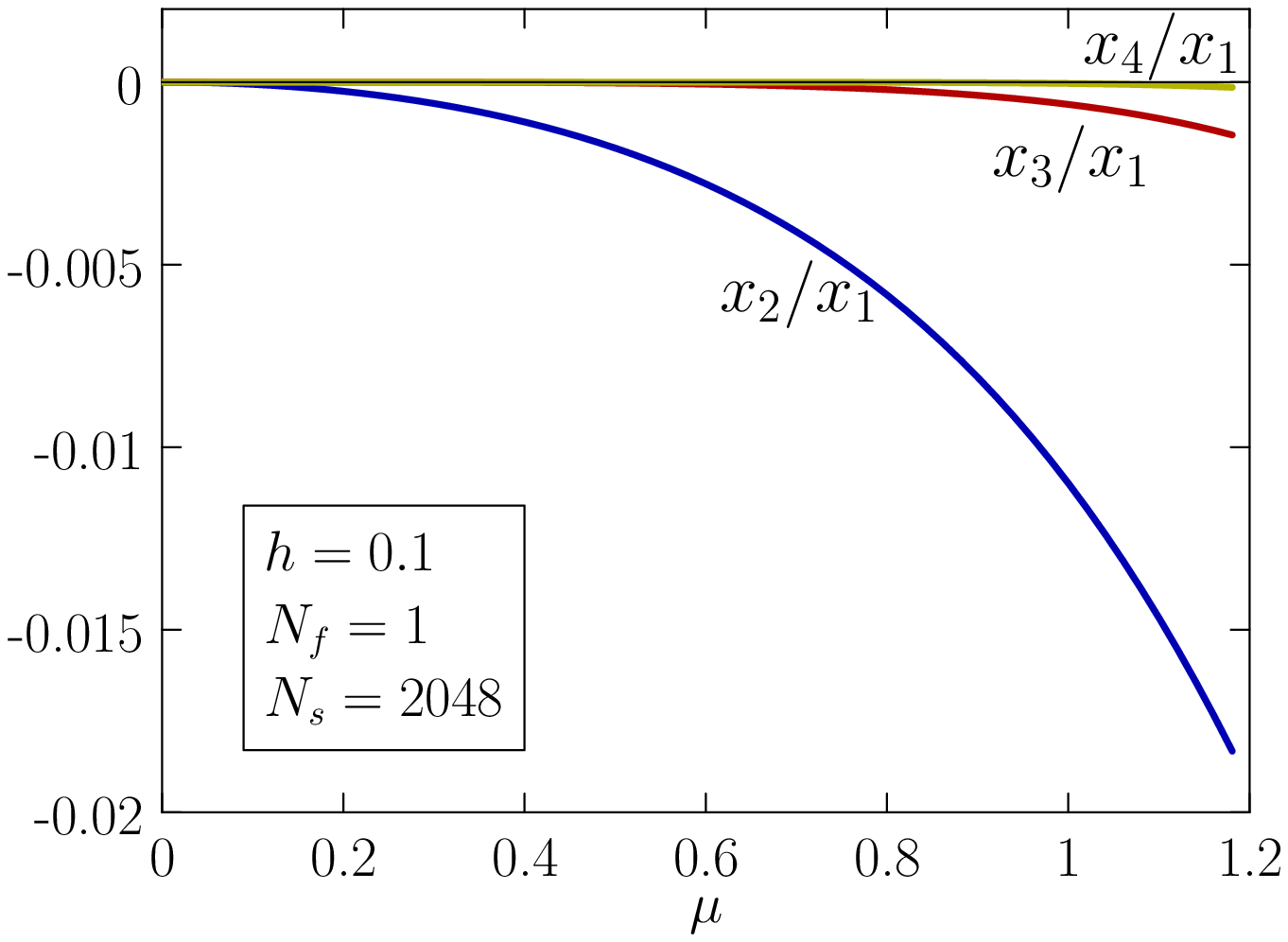}
\end{minipage}
\caption{At ${\cal O}(\lambda_1^0)$ in the lattice strong coupling and hopping expansion: (left) $x_n$ for $n = 1, ..., 4$ as a function of $\mu/T$, (right) $x_2/x_1, ..., x_4/x_1$ as a function of $\mu/T$.}
\label{xn-plots}
\end{figure}

Even though our results indicate that the higher order cumulants, or $x_n$ with $n > 1$, are non-zero at strong coupling, $\lambda_1 \rightarrow 0$, they are small compared to $x_1$ in the regime of validity of the hopping expansion, $h e^{\mu/T} \ll 1$. In Figure \ref{xn-plots} (left) we plot $x_n$ for $n = 1, ..., 4$ as a function of $\mu/T$. In Figure \ref{xn-plots} (right) we plot the ratios $x_2/x_1, x_3/x_1, x_4/x_1$ as a function of $\mu/T$. The plots include all contributions of $x_1$ and $x_2$, up to ${\cal O}(h^6)$ at ${\cal O}(\lambda_1^0)$, and the leading order contributions to $x_3, ..., x_6$. The results are calculated for values of $h$ and $\mu$ which are towards the border of the region of validity so they should be interpreted with caution, but they do indicate that in the region of strong coupling, large quark masses, and small chemical potentials, the higher order cumulants represented by the $x_n$ for $n > 1$ are small compared compared to $x_1$. This is consistent with the recent simulation results in \cite{Saito:2013vja}. Whether or not the higher order cumulants are ever significant compared to $x_1$ is a question that will need to be addressed non-perturbatively.

\section{Distribution of the complex phase}

It is worthwhile at this point to make a few comments regarding the how to best check the validity of the Gaussian approximation. The cumulants are all ${\cal O}(V)$, so it is possible that they could be comparable in some regions of the phase diagram. However, even if they are comparable, it is not necessarily the case that the distribution of the complex phase would look noticeably different from a Gaussian.

\begin{figure}
\begin{minipage}{0.48\textwidth}
\includegraphics[width=0.96\textwidth]{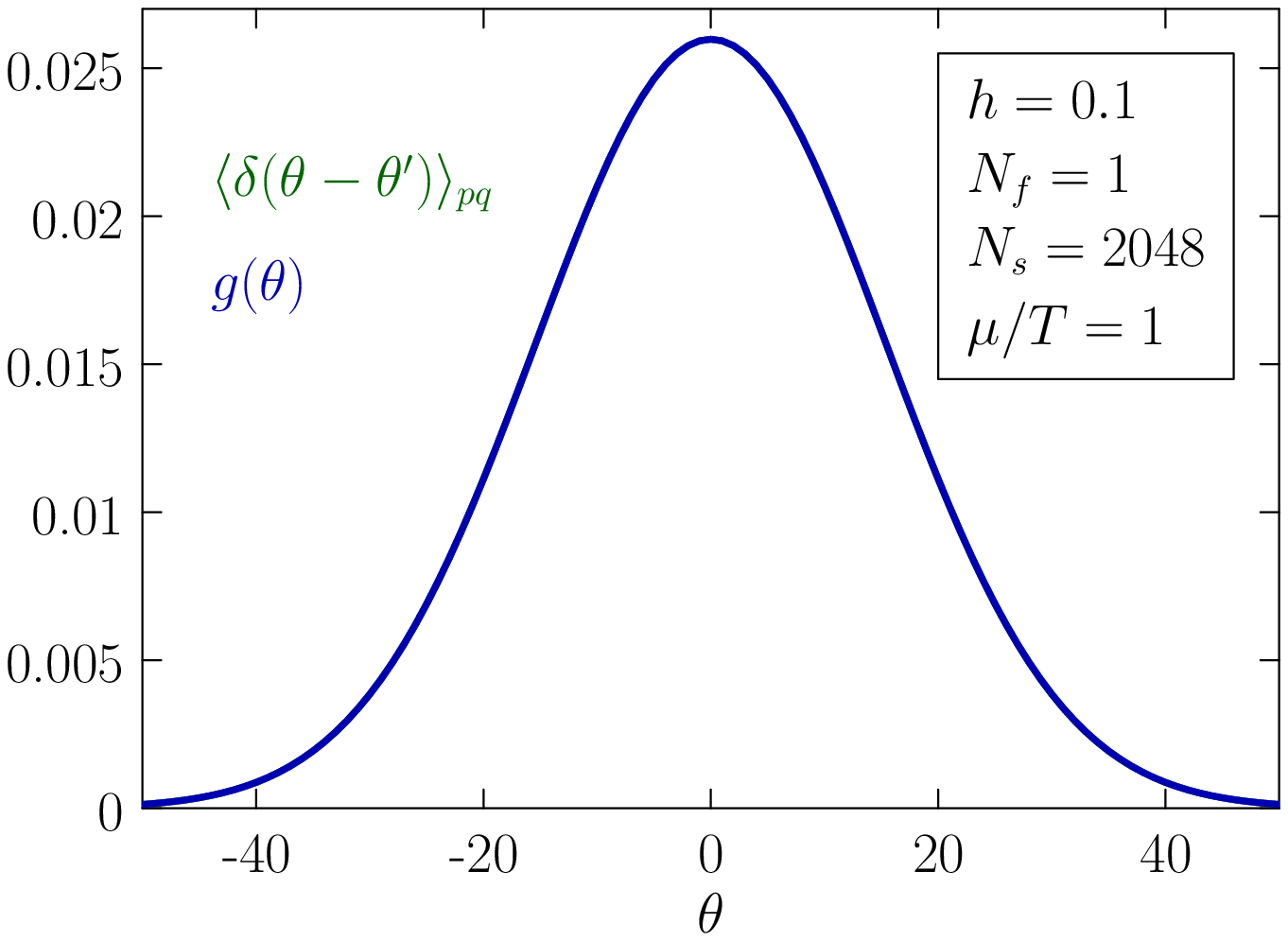}
\end{minipage}
\begin{minipage}{0.48\textwidth}
\includegraphics[width=0.96\textwidth]{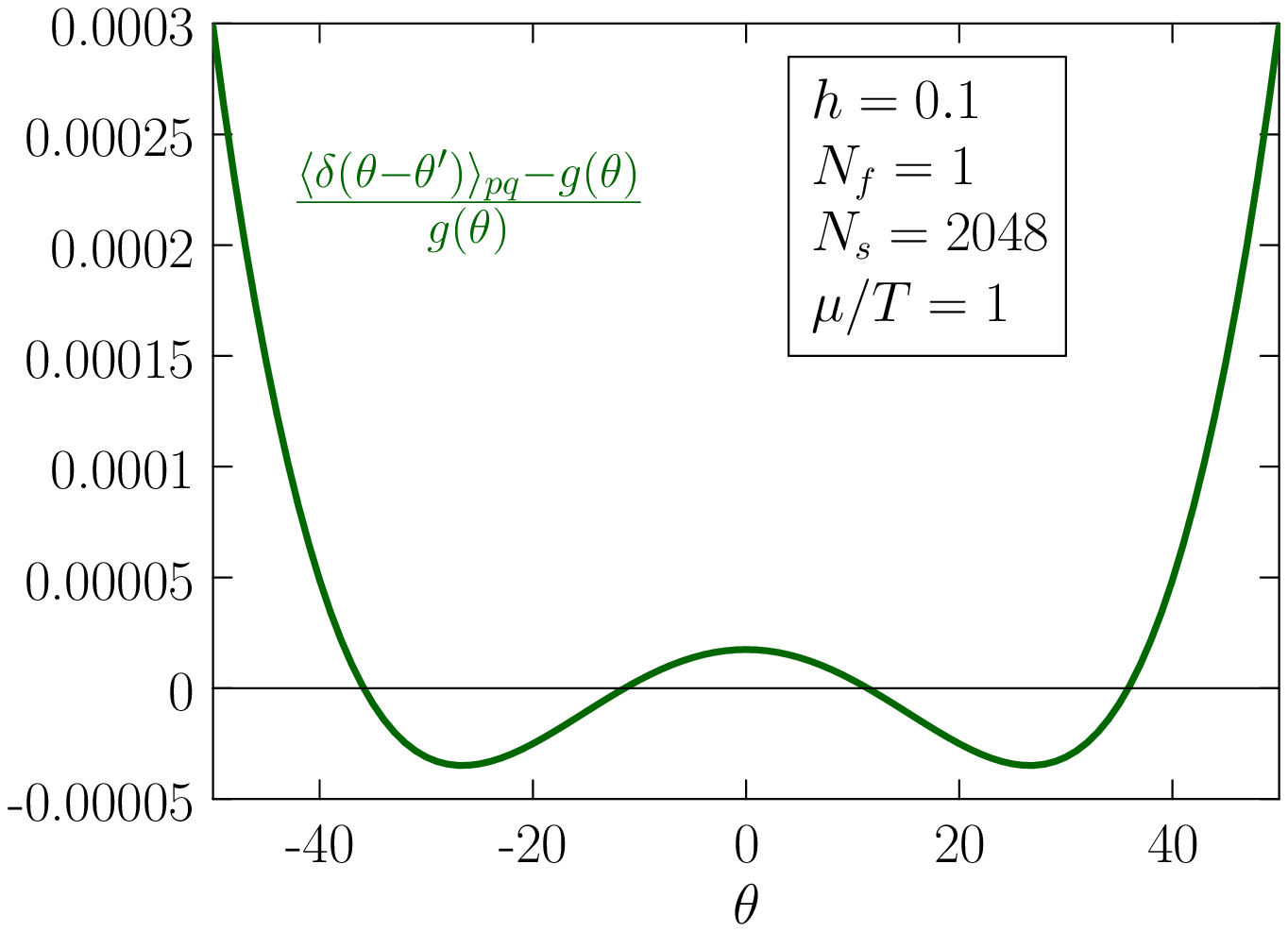}
\end{minipage}
\caption{At $\mu/T = 1.0$: (left) Distribution of the complex phase $\rho_{pq}(\theta)$ (blue), compared with a Gaussian distribution $g(\theta)$ (green), as a function of $\theta$, (right) fractional difference in the distributions $\frac{\rho_{pq}(\theta) - g(\theta)}{g(\theta)}$.}
\label{data-dist}
\end{figure}

Using our results for the $x_n$ from the strong coupling and hopping expansions it is possible to calculate the distribution $\rho_{pq}(\theta)$ which takes the form in (\ref{gen-dist}), and compare with a Gaussian form $g(\theta) = \frac{1}{\sqrt{\pi x_1}} e^{-\theta^2/x_1}$. The results are plotted in Figure \ref{data-dist} (left), which are obtained for $h=0.1$, $N_f = 1$, $N_s = 2048$, $\mu/T = 1$, where the hopping expansion is approaching its edge of validity $h e^{\mu/T} \ll 1$ (as $h e^{\mu/T}$ is decreased $x_2$, $x_3$, ... become less significant compared to $x_1$). Since the difference between the actual distribution and the Gaussian form are indistinguishable by eye it is helpful to consider the fractional difference
\EQ{
\frac{\rho_{pq}(\theta) - g(\theta)}{g(\theta)} \, .
}
This is plotted in Figure \ref{data-dist} (right), which shows that the corrections are ${\cal O}\left( \frac{1}{V} \right)$. It is possible to see this analytically by Taylor expanding the exponentials in $x_2$, $x_3$, ... in the moments $\langle e^{2 i p \theta'} \rangle$ in (\ref{gen-moms}) in order to calculate the integral over $p$ in the distribution (\ref{gen-dist})
\SP{
\rho_{pq}(\theta) &\sim \frac{1}{\pi} \sum_{k=0}^{\infty} \frac{(- x_2)^k}{k!} \sum_{l=0}^{\infty} \frac{(- x_3)^l}{l!} \times ... \int_{-\infty}^{\infty} {\rm d}p~ p^{4k+6l+...} e^{-2 i p \theta} e^{-p^2 x_1} \, ,
}
where
\EQ{
\frac{1}{\pi} \int_{-\infty}^{\infty} {\rm d}q~ q^{\alpha} e^{-2 i q \theta} e^{-q^2 x_1} = \frac{1}{\pi x_1^{(\alpha+1)/2}} \Gamma\left( \frac{\alpha+1}{2} \right) {}_{1}F_{1} \left( \frac{\alpha+1}{2} ; \frac{1}{2} ; -\frac{\theta^2}{x_1} \right) \, ,
}
with $\alpha \ge 0$ and even, and $x_1 > 0$. This integral goes to the Gaussian form as $\alpha \rightarrow 0$. Expanding in powers of $x_2$, $x_3$, $...$, one obtains
\EQ{
\rho_{pq}(\theta) = \frac{1}{\sqrt{\pi x_1}} e^{-\theta^2/x_1} \left[ 1 - \frac{3 x_2}{4 x_1^2} + \frac{3 x_2 \theta^2 - \frac{15}{8} x_3}{x_1^3} + ... \right] \, .
}
Since $x_2$, $x_3$, ... are ${\cal O}(V)$, this result is an expansion around the Gaussian form in powers of $\frac{1}{V}$.

It is worthwhile to clarify that our results are consistent with the central limit theorem, which is an argument in support of the Gaussian approximation from probability theory. The central limit theorem states that the the distribution of a collection of independent data points approaches a Gaussian form in the limit of a large enough sample size. For the case at hand the limit where each measurement of the complex phase phase angle becomes independent corresponds to the infinite volume limit, such that the volume is much larger than the correlation length. The corrections we have found to a Gaussian form of the distribution begin at ${\cal O}\left( \frac{1}{V} \right)$. Nevertheless, the corrections contribute at leading order to $\langle e^{i N_f \theta} \rangle$, that is, the central limit theorem is not a sufficient reason to use only the first cumulant.

\begin{figure}
\begin{minipage}{0.48\textwidth}
\includegraphics[width=0.96\textwidth]{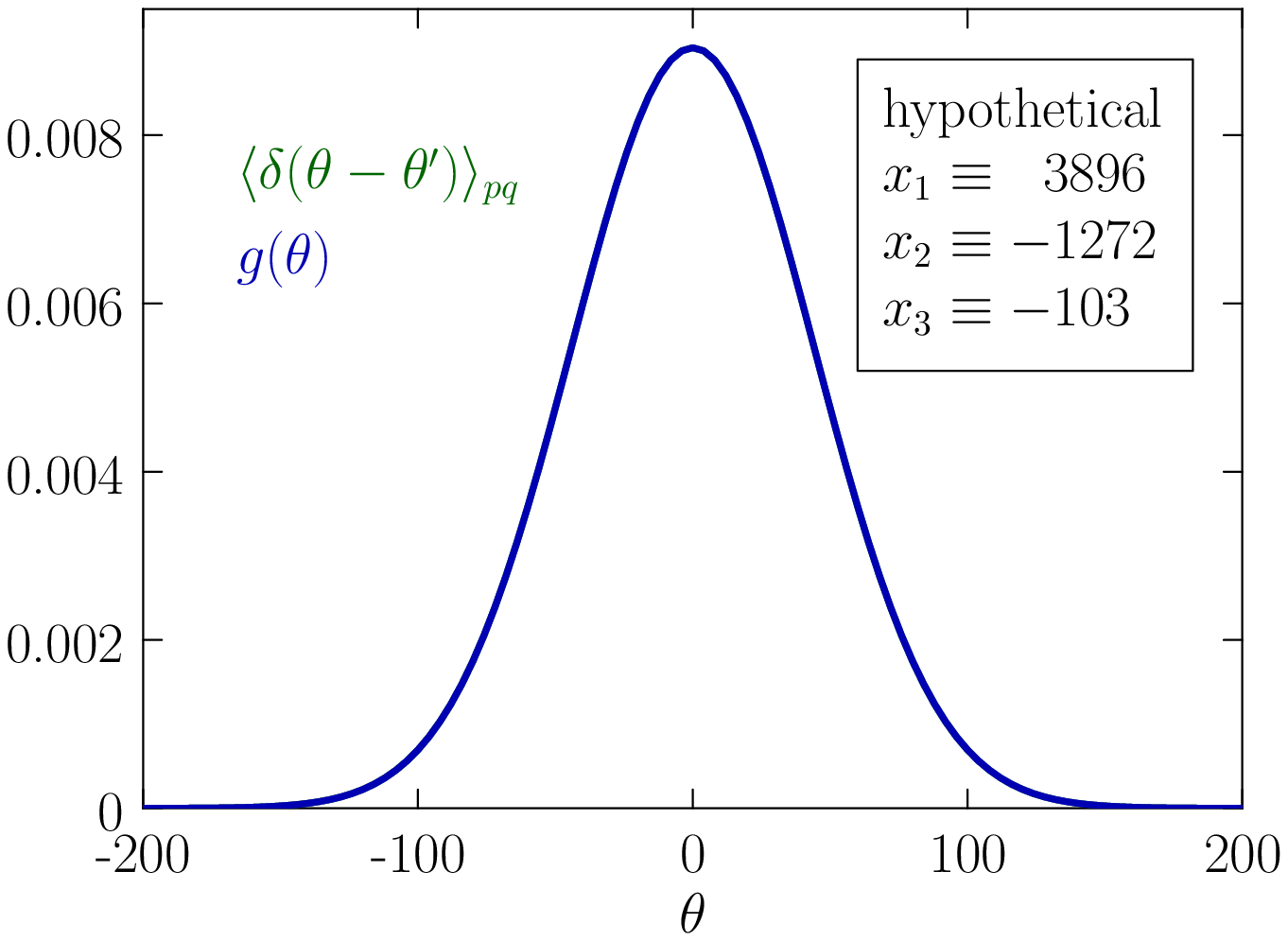}
\end{minipage}
\begin{minipage}{0.48\textwidth}
\includegraphics[width=0.96\textwidth]{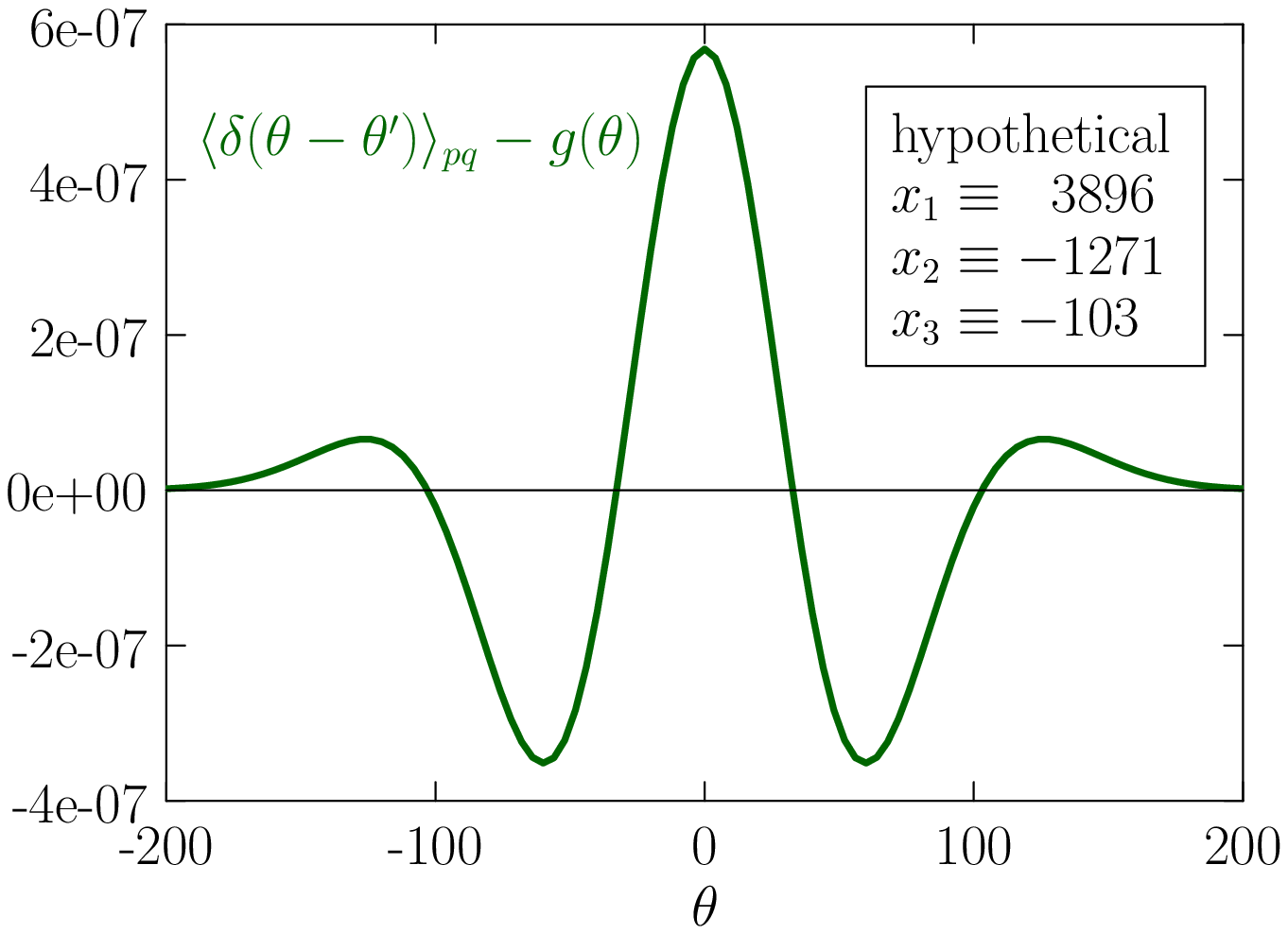}
\end{minipage}
\caption{At $\mu/T = 1.0$: (left) Distribution of the complex phase $\rho_{pq}(\theta)$ (blue), compared with a Gaussian distribution $g(\theta)$ (green), as a function of $\theta$, (right) Difference in distributions $\rho_{pq}(\theta) - g(\theta)$.}
\label{hypo-dist}
\end{figure}

At this point one might argue that the distribution appears to take an almost Gaussian form because we are working in a region of phase diagram where $x_2, x_3, ...$ are much smaller than $x_1$. However, it is possible to demonstrate that the distribution of the complex phase can also be indistinguishable from a Gaussian, and consistent with the central limit theorem, even when the cumulants are comparable in magnitude (see also \cite{Greensite:2013gya}). To make this point we choose hypothetical values of $x_1, x_2, x_3$ which are more comparable, and re-plot the distribution along with the Gaussian form. This is shown in Figure \ref{hypo-dist} (left). Note that we kept $x_4, x_5, x_6$ as before where $x_6 > 0$ ensures convergence.

In Figure \ref{hypo-dist} (right) we plot the ordinary difference in the distributions $\rho_{pq}(\theta) - g(\theta)$, which shows that the corrections to a Gaussian form in $\rho_{pq}(\theta)$ appear in the central region of the distribution, and that the corrections are sufficiently small that even significant contributions from $x_2$, $x_3$, ... could be impossible to see by considering the shape of the distribution alone.

\section{Binder cumulant}

Another method of measuring the validity of the Gaussian distribution is offered by calculating Binder cumulants. For example, the first relevant Binder cumulant is
\EQ{
B_4^{\theta} \equiv \frac{\langle \theta^4 \rangle}{\langle \theta^2 \rangle^2} \, .
}
Since the fourth cumulant of the expansion of $\langle e^{i N_f \theta} \rangle$ is defined by
\EQ{
\langle \theta^4 \rangle_c = \langle \theta^4 \rangle - 3 \langle \theta^2 \rangle^2 \, ,
}
this indicates that
\EQ{
B_4^{\theta} \rightarrow 3 ~~~\text{as}~~~ \langle \theta^4 \rangle_c \rightarrow 0 \, .
}
Simulation results of Ejiri in \cite{Ejiri:2007ga} with $\frac{m_{\pi}}{m_{\rho}} \approx 0.7$, using the improved staggered quark action and working on $16^3 \times 4$ lattices, indicate that $B_4^{\theta}$ is close to $3$. However, it is important to notice that $B_4^{\theta}$ is expected go to $3$ in the large volume limit. This is because $\langle \theta^4 \rangle_c = (B_4^{\theta} - 3) \langle \theta^2 \rangle^2 = {\cal O}(V)$, and $\langle \theta^2 \rangle^2 = {\cal O}(V^2)$. Therefore
\EQ{
B_4^{\theta} - 3 = {\cal O} \left( \frac{1}{V} \right) \, .
}
Since this result is only based on the fact that the cumulants are all ${\cal O}(V)$ the message here is that it could well be that $B_4^{\theta} \rightarrow 3$, even while the $x_n$, for $n > 1$, are significant. It would be interesting to see how the $x_n$ compare using the simulation parameters in \cite{Ejiri:2007ga}.

\section{Discussion}

It is constructive to discuss the simulation results which exist already that have made use of the Gaussian approximation. These results are obtained by the WHOT-QCD collaboration. In \cite{Saito:2012nt,Saito:2013vja} WHOT-QCD presents simulation results in the heavy quark limit using the unimproved Wilson quark action, and $24^3 \times 4$ lattices. Their results for the distribution indicate that it takes a Gaussian form for a wide range of values of the average Polyakov line. A comparison of the leading and higher order cumulants indicates that the second cumulant $\langle \theta^2 \rangle_c$ is always dominant. In terms of the hopping parameter and chemical potential, data taken for $\kappa^4 \sinh(\mu/T) = 0.00002$ in \cite{Saito:2013vja} (see Figure 9) indicate that the higher order cumulants appear to be consistent with zero, but for $\kappa^4 \sinh(\mu/T) = 0.00005$, $\langle \theta^4 \rangle_c$ has grown compared to $\langle \theta^2 \rangle_{c}$, in particular when the Polyakov line approaches zero. This is consistent with our results in that there is increased importance of the higher order cumulants as the chemical potential is increased, or as the quark mass is decreased.

What is perhaps more surprising is that simulation results for light quarks \cite{Nakagawa:2011eu} with $\frac{m_{\pi}}{m_{q}} \approx 0.8$, using the improved Wilson quark action and working on smaller lattices $8^3 \times 4$, also indicate that the distribution takes a Gaussian form, even at large values of the chemical potential $\mu/T = 0.4, 2.4$ (see Figure 3 in \cite{Nakagawa:2011eu}). At $\mu/T = 0.4$ the higher order cumulant $x_4 = \frac{1}{4!} \langle \theta^4 \rangle_c$ appears to be consistent with zero (see Figure 4 in \cite{Nakagawa:2011eu}), but at $\mu/T = 1.2$ it is difficult to judge due to error bars which are sufficiently large that $x_4$ could be comparable to $x_2$, so it will be interesting to see what new results will show.

It is also important to understand why our results for the distribution $\rho_{pq}(\theta)$ from the strong coupling and hopping expansion differs from that of the hadron resonance gas model. In \cite{Langelage:2010yn} the authors calculate the pressure from the strong coupling and hopping expansion and find that the result matches on precisely to that from the hadron resonance gas model. There is no inconsistency. The calculation in \cite{Langelage:2010yn} is performed including terms up to ${\cal O}(h^3)$ in the hopping expansion. In our calculation the differences in the distributions only start to appear at ${\cal O}(h^4)$. To obtain a contribution at this order from the hadron resonance gas model, which would lead to a nonzero $x_2$, one would need to consider bound states of at least $4$ quarks, since that would be the only way to obtain a contributions at ${\cal O}(p^4)$ in $\log \langle e^{2 i p \theta'} \rangle$. However, since $4$ quarks can not combine to give nonzero contributions for $N_c = 3$ which vanish at $N_c = \infty$, we expect that one would actually need to consider bound states of at least $6$ quarks.

\section{Conclusions}

We have calculated the leading order contributions to the the first six cumulants in a cumulant expansion of the complex phase $\langle e^{i N_f \theta} \rangle_{pq}$, using the hadron resonance gas model, and a combined lattice strong coupling and hopping parameter expansion. Considering free ground state mesons of spin $0$ and $1$, and baryons of spin $\frac{1}{2}$ and $\frac{3}{2}$, we find that the distribution of the complex phase takes a perfectly Gaussian form. However, when the strong coupling and hopping expansion are considered there are corrections which begin to appear at ${\cal O}(h^4)$ for ${\cal O}(\lambda_1)$ and at ${\cal O}(h^6)$ for ${\cal O}(\lambda_1^0)$. These appear to grow as the quark mass and coupling strength decrease, or as the chemical potential increases.


The main implication of our work is that in order to justify truncating the cumulant expansion to the second order cumulant, it is is necessary to show that higher order cumulants are negligible.  For this purpose, neither the apparent Gaussianity of the phase angle distribution, nor the near agreement of the Binder cumulant $B_4^{\theta}$ with $3$ is sufficient.  In either case, corrections on the order of $\frac{1}{V}$ or smaller can be associated with significant higher-order cumulants. Thus measurements of the phase angle moment $\langle \theta^4 \rangle_c$ to an accuracy of at least ${\cal O}(\frac{1}{V})$ is required.  For higher order cumulants, the phase angle moments would have to be computed to accuracies of even higher powers of $\frac{1}{V}$.

\section{Acknowledgements}

We would like to thanks the organizers of Lattice 2013 for the chance to present this work in the plenary session. We would like to give special thanks to Hana Saito and Shinji Ejiri for helpful discussions. This research was supported by the U.S. Department of Energy grant DE-FG03-92ER40711 (JG) and the {\it Sapere Aude} program of the Danish Council for Independent Research (JCM and KS).


\end{document}